\documentclass[prc,twocolumn,superscriptaddress,showpacs,twoside,floatfix]
{revtex4}

\usepackage{amssymb,epsfig}

\begin{document}

\title{New insight into the low-energy $^9$He spectrum}

\author{M.S.~Golovkov}
\author{L.V.~Grigorenko}
\author{A.S.~Fomichev}
\author{A.V.~Gorshkov}
\author{V.A.~Gorshkov}
\author{S.A.~Krupko}
\author{Yu.Ts.~Oganessian}
\author{A.M.~Rodin}
\author{S.I.~Sidorchuk}
\author{R.S.~Slepnev}
\author{S.V.~Stepantsov}
\author{G.M.~Ter-Akopian}
\affiliation{Flerov Laboratory of Nuclear Reactions, JINR, Dubna, RU-141980
Russia}
\author{R.~Wolski}
\affiliation{Flerov Laboratory of Nuclear Reactions, JINR, Dubna, RU-141980
Russia}
\affiliation{The Henryk Niewodnicza\'nski Institute of Nuclear
Physics, Krak\'{o}w, Poland}
\author{A.A.~Korsheninnikov}
\author{E.Yu.~Nikolskii}
\affiliation{RIKEN, Hirosawa 2-1, Wako, Saitama 351-0198, Japan}
\affiliation{RRC ``The Kurchatov Institute'', Kurchatov sq.\ 1, 123182
Moscow, Russia}
\author{V.A.~Kuzmin}
\author{B.G.~Novatskii}
\author{D.N.~Stepanov}
\affiliation{RRC ``The Kurchatov Institute'', Kurchatov sq.\ 1, 123182
Moscow, Russia}
\author{S.~Fortier}
\affiliation{Institut de Physique Nucleaire, IN2P3-CNRS, F-91406 Orsay, France}
\author{P.~Roussel-Chomaz}
\author{W.~Mittig}
\affiliation{GANIL, BP 5027, F-14076 Caen Cedex 5, France}

\date{\today. {\tt File: he9-11.tex }}

\begin{abstract}
The spectrum of $^9$He was studied by means of the
$^8$He($d$,$p$)$^9$He reaction at a lab energy of 25 MeV/n and
small center of mass (c.m.) angles. Energy and angular
correlations were obtained for the $^9$He decay products by
complete kinematical reconstruction. The data do not show narrow
states at $\sim $1.3 and $\sim $2.4 MeV reported before for
$^9$He. The lowest resonant state of $^9$He is found at about 2
MeV with a width of $\sim $2 MeV and is identified as $1/2^-$. The
observed angular correlation pattern is uniquely explained by the
interference of the $1/2^-$ resonance with a virtual state $1/2^+$
(limit on the scattering length is obtained as $a
> -20$ fm), and with the $5/2^+$ resonance at energy $\geq 4.2$ MeV.
\end{abstract}


\maketitle


\textit{Introduction.} --- Since the first observation of $^9$He in the
experiment \cite{set87} it was studied in relatively small number of works
compared to the neighbouring exotic neutron dripline nuclei. This can be
connected, on one hand, to the facts that technical difficulty of the
precision measurements rapidly grows with move away from the stability line. On
the other hand, already in the first experiment (pion double charge exchange on
the $^9$Be nucleus \cite{set87}) several narrow resonances were observed above
the $^8$He+$n$ threshold. This observation was confirmed in Ref.\ \cite{boh99},
where the $^9$Be($^{14}$C,$^{14}$O)$^9$He reaction was used, and now the
experimental situation with the low-energy spectrum of $^9$He is considered to
be well established. A new rise of interest to $^9$He was connected with the
question of a possible $2s$ state location in the framework of the shell
inversion
problem in nuclei with large neutron excess. The recent experiment \cite{che01}
was focused on the search for the virtual state in $^9$He. An upper limit on the
scattering length $a<-10$ fm was established in this work. The properties of
states in $^{9}$He were inferred in \cite{gol03} basing on studies of isobaric
partners in $^{9}$Li. The available results are summarized in Table
\ref{tab:exp}.

Interpretation of the $^9$He spectrum as provided in
\cite{set87,boh99} faces certain difficulties which were not
unnoticed (e.g.\ Ref.\ \cite{bar04}). Indeed, the ground $1/2^-$
state is expected to be single particle state with width estimated
as $0.8-1.3$ MeV at $E = 1.27$ MeV for typical channel radii $3 -
6$ fm. This requires spectroscopic factor $S\sim 0.1$ which
contradicts single particle character of the state. F.\ Barker in
Ref.\ \cite{bar04} concludes on this point that ``some configuration mixing
in either the $^{9}$He($1/2^-$) or $^{8}$He($0^+$) state or both
is possible, but is unlikely to be large enough to reduce the
calculated width to the experimental value''. The next, presumably
$3/2^-$ state, should be a complicated particle-hole excitation as
$p_{3/2}$ subshell is occupied. However, a much larger
spectroscopic factor $S\sim 0.3-0.4$ is required for its widths
found in a range $2.0-2.6$ MeV.

\begin{figure}
\includegraphics[width=0.44\textwidth]{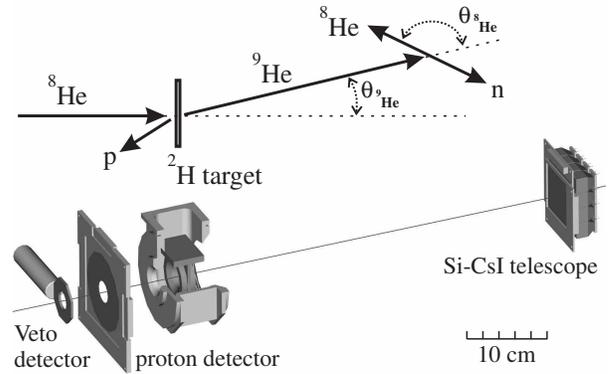}
\caption{Experimental setup, angles, and momenta.}
\label{fig:setup}
\end{figure}

Having in mind the mentioned problematic issues we decided to
study the $^9$He in the ``classical'' one-neutron transfer
($d$,$p$) reaction well populating single particle states. In
contrast with the previous works complete kinematics studies were
foreseen to reveal the low-energy $s$-wave mode. Following the experimental
concept of \cite{gol04a,gol05b}, where correlation
studies of $^5$H continuum were accomplished by means of the
$^3$H($t$,$p$) transfer reaction, this work was performed in the
so called ``zero geometry''.

\begin{table}[b]
\caption{Experimental positions of states in $^9$He relative to the $^8$He+$n$
threshold (energies and widths are given in MeV).}
\begin{ruledtabular}
\begin{tabular}[c]{cccccccc}
  &  1/2$^+$  & \multicolumn{2}{c}{1/2$^-$}
  &  \multicolumn{2}{c}{3/2$^-$}  &  \multicolumn{2}{c}{5/2$^+$}   \\
Ref.   & $a$ (fm)  & $E$ & $\Gamma$ & $E$ & $\Gamma$ & $E$ & $\Gamma$  \\
\hline
\cite{set87}  &       & 1.13(10) & small\footnotemark[2]   & 2.3 &
small\footnotemark[2] & 4.9
&    \\
\cite{boh99}  &       & 1.27(10) & 0.1(0.6) & 2.42(10) & 0.7(2) & 4.3 & small
\\
\cite{che01}  & $< \! -10$ &      &  &  &   &    &   \\
\cite{gol03}\footnotemark[1]
              &    & 1.1  &   & 2.2 &     & 4.0 &       \\
Our           & $> \! -20$ &  2.0(0.2)  &  2 &    &    & $\geq 4.2$ & $> 1$   \\
\end{tabular}
\end{ruledtabular}
\label{tab:exp}
\footnotetext[1]{Inferred from isobaric symmetry.}
\footnotetext[2]{Observed peak width is smaller than the declared resolution.}
\end{table}


\begin{figure*}
\includegraphics[width=0.86\textwidth]{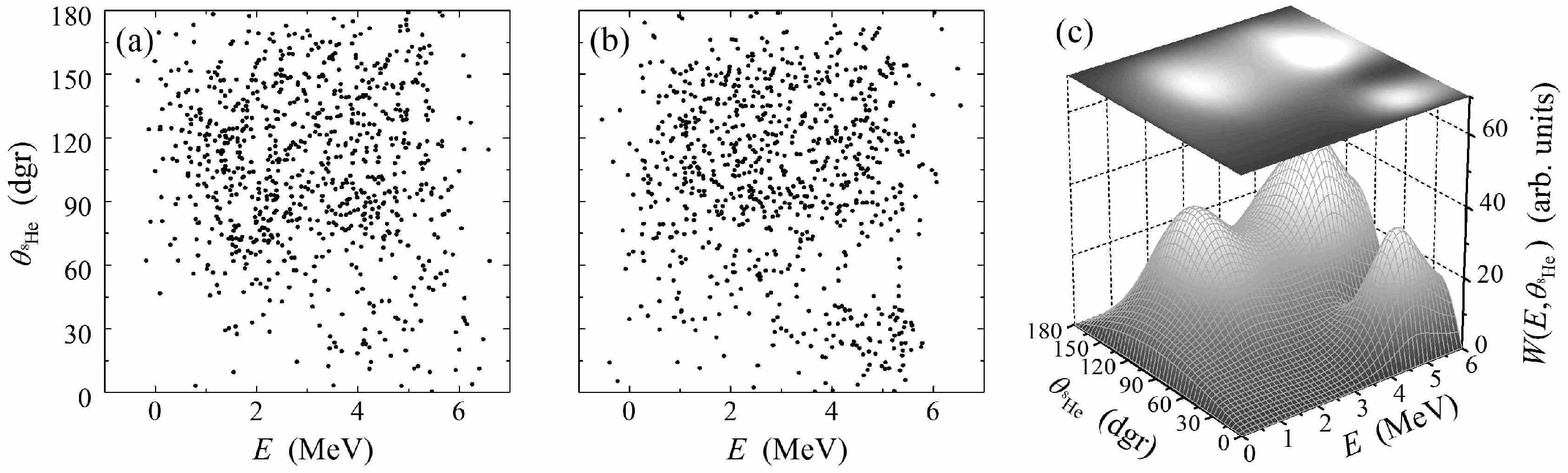}
\caption{(a) Experimental data on the $\{E,\theta_{^8\text{He}}
\}$ plane. (b) MC simulation [with the same statistics as (a)]
using the input shown in panel (c). (c) Theoretically
reconstructed distribution with parameter set 1 from Table
\ref{tab:th}. Hereafter, the $^9$He energy $E$ (missing mass) is
given relative to the $^8$He+$n$ breakup threshold.}
\label{fig:expth}
\end{figure*}

\textit{Experiment.} --- The experiment was done at the U-400M
cyclotron of the Flerov Laboratory of Nuclear Reactions, JINR
(Dubna, Russia). A 34 MeV/nucleon $^{11}$B primary beam delivered
by the cyclotron hitted a 370 mg/cm$^2$ Be production target. The
modified  ACCULINNA fragment separator \cite{rod97}  was used to
produce a $^8$He secondary beam  with a typical intensity of
$2\times 10^4$ s$^{-1}$. The beam was focused on  a cryogenic
target \cite{yuk03} filled with deuterium at 1020 mPa pressure and
cooled down to 25~K. The 4 mm thick target cell was supplied with
6 $\mu$m stainless steel windows, 30 mm in diameter.

Experimental setup and kinematical diagram for the
$^{2}$H($^8$He,$p$)$^9$He reaction are  shown in Fig.\
\ref{fig:setup}. Slow protons escaping from the target in the
backward direction hitted on an annular 300 $\mu$m silicon
detector with an active area of the outer and inner diameters of
82 mm and 32 mm, respectively, and a 28 mm central hole. The
detector was installed 100 mm upstream of the target. It was
segmented in 16 rings on one side and 16 sectors on the other side
providing a good position resolution. The detection threshold for
the protons ($\sim $1.2 MeV) corresponded to a $\sim$5.5 MeV
cutoff in the missing mass of $^9$He. We did not use here particle
identification because, due to the kinematical constraints of the
$^8$He+$^2$H collisions, only protons can be emitted in the
backward direction. The main cause of the background was due to
evaporation protons originating from the interaction of $^8$He
beam with the material of target windows. This background was
almost completely suppressed by the coincidence with $^8$He. The
detection of such coincidences fixed the complete kinematics for
the experiment. Energy-momentum conservation was used for cleaning
the spectra. Finally the comparison with an empty target run has
shown that only $\sim 2 \%$ events can be treated as a background.

The $^8$He nuclei resulting from the $^9$He decay, focused in
narrow angular cone relative to the beam direction, were detected
by a Si-CsI telescope mounted in air just behind the exit window
of the scattering chamber. The 82 mm diameter exit window was
closed by a 125 $\mu$m capton foil. The Si-CsI telescope consisted
of two 1 mm thick silicon detectors and 16 CsI crystals with
photodiode readouts. The $6 \times 6$ cm Si detectors were
segmented in 32 strips both in horizontal and vertical directions,
providing position resolution and particle identification by the
$\Delta E$-$E$ method (together with following thick CsI
detector). Sixteen $1.5 \times 2 \times 2$ cm CsI crystals were
arranged as a $4 \times 4$ wall just behind the Si detectors. A $
2 \%$ energy resolution of the CsI detectors allowed particle
identification even for $Z=1$ nuclei. The distance between the
target and the telescope (50 cm) was sufficient to provide a good
efficiency for the detection of the $^8$He nuclei in coincidence
with protons in the whole range of accessible $^9$He energies. To
eliminate signals in the proton telescope coming from the beam
halo a veto detector was installed upstream the proton telescope.
The energy of the $^8$He beam in the middle of the target was
$\sim $25 MeV/nucleon. Energy spread of the beam, angular
divergence, and position spread on the target were $8.5 \%$,
$0.23^{\circ}$, and about 5.4 mm respectively. A set of beam
detectors was installed upstream of the veto (not shown in the
Fig.\ \ref{fig:setup}). The beam energy was measured by two
time-of-flight plastic scintillators with a 785 cm base. The
overall time resolution was 0.8 ns. Beam tracking was made by two
multiwire proportional  chambers installed 26 and 80 cm upstream
of the target. Each chamber had two perpendicular planes of wires
with a 1.25 mm pitch. Energy resolution was estimated by
Monte-Carlo (MC) code taking into account all experimental
details. It was found to be 0.8 MeV (FWHM) for the $^9$He missing
mass.

\begin{figure*}
\includegraphics[width=0.90\textwidth]{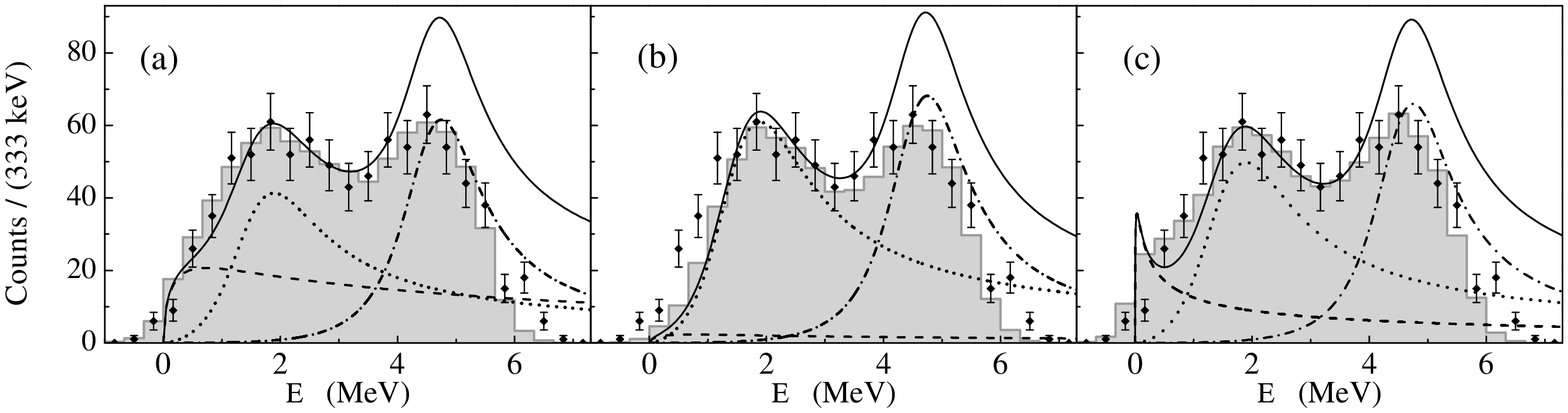} \\
\includegraphics[width=0.85\textwidth]{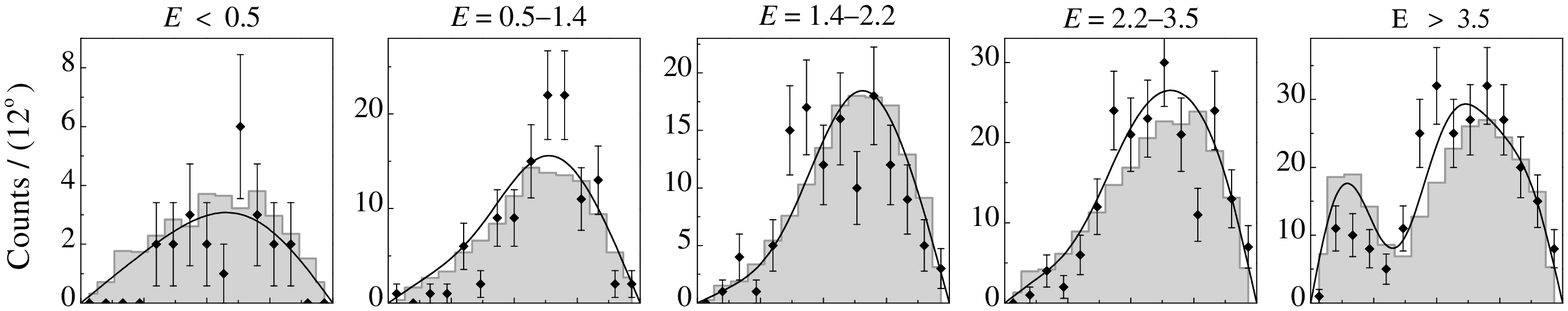}\\
\includegraphics[width=0.85\textwidth]{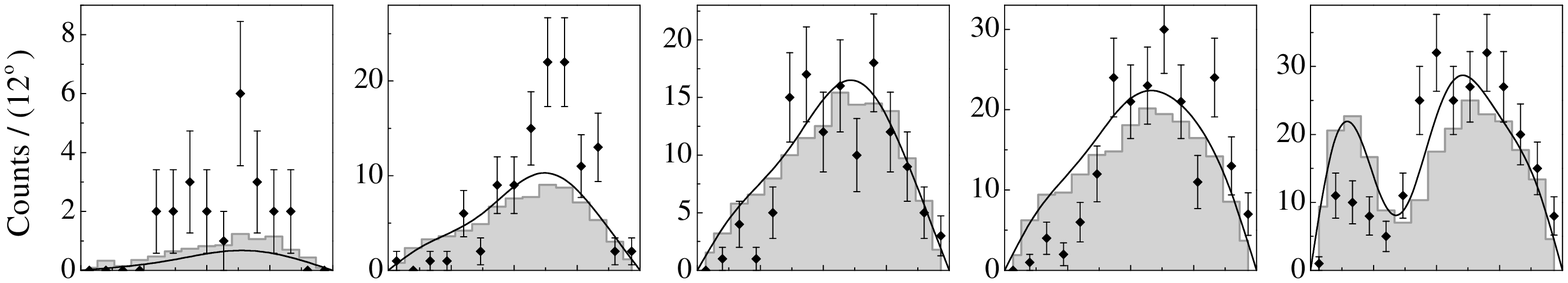}\\
\includegraphics[width=0.85\textwidth]{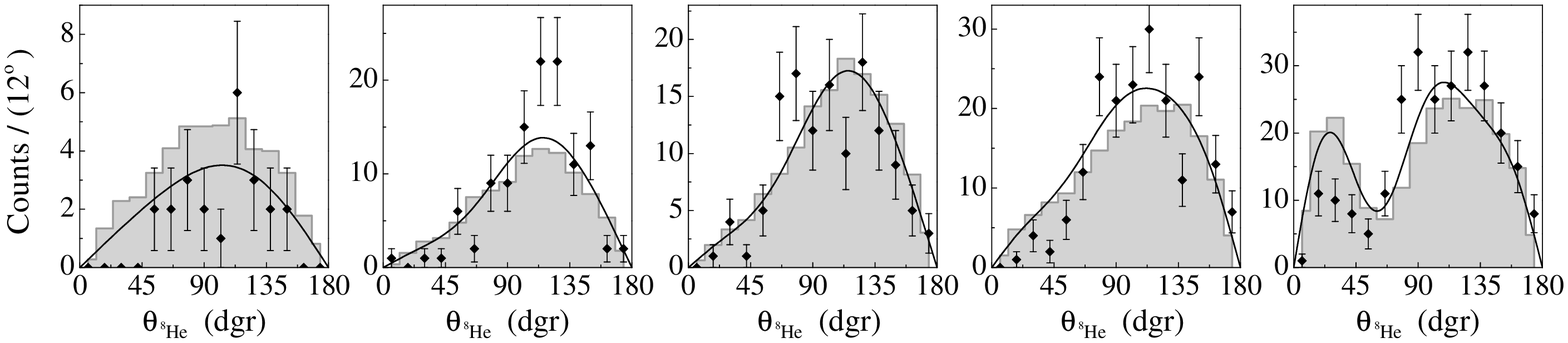}
\caption{Inclusive energy spectrum for the data of Fig.\
\ref{fig:expth}a is shown by diamonds. Theoretical curves in
panels (a), (b) and (c) correspond to parameter sets 1, 2, and 3
from Table \ref{tab:th}. Solid line is total result, while dashed,
dotted and dash-dotted curves stand for contributions of $1/2^+$,
$1/2^-$, and $5/2^+$ states. Gray histogram shows the results of
MC simulation. The angular distributions for certain energy bins
corresponding to panels (a), (b), and (c) are show in second,
third and fourth rows.}
\label{fig:distrib}
\end{figure*}


\textit{Qualitative considerations.} --- Data obtained in the
experiment are shown in Figs.\ \ref{fig:expth} and
\ref{fig:distrib}. The total number of counts corresponds to cross
section of $\sim$7 mb/sr. This value is consistent with the direct
one-neutron transfer reaction mechanism at forward angles.

The narrow states known from the literature do not show up in the
data. Instead, we get two broad peaks at about 2 and 4.5 MeV. Near
the threshold the $^9$He spectrum exhibits behavior (although
distorted by the finite energy resolution of the experiment) which
is more consistent with $s$-wave ($\sigma \sim \sqrt{E}$) rather
than $p$-wave  ($\sigma \sim E^{3/2}$). This is an indication for
a possible virtual state in $^9$He.

An important feature of the data is a prominent forward-backward
asymmetry with $^8$He flying preferably in the backward direction
in the $^{9}$He c.m.\ system. This is not feasible if the
narrow (means long-living) states of the $^9$He are formed. To
describe such an asymmetry the interference of opposite parity
states is unavoidable. As far as asymmetry is observed even at
very low energy, the $s$-$p$ interference is compulsory. Such an
interference can provide only a very smooth distribution described
by the first order polynomial [Eq.\ (\ref{eq:sigma-full}) and
Fig.\ \ref{fig:distrib}, $E<2.2$ MeV]. Since above 3 MeV the
character of the distribution changes to  a higher polynomial, but
asymmetry does not disappear, the $p$-$d$ interference is also
needed. This defines the minimal set of states as $s$, $p$, and
$d$.


\textit{Theoretical model.} --- In the zero geometry approach the
resonant states of interest are identified by the observation of
recoil particle (here proton) at zero (in reality small,
$3^{\circ} \leq \theta_{p}(\text{c.m.})\leq 7^{\circ}$) angle.
This means that the angular momentum transferred to the studied
system should have zero projection on the axis of the momentum
transfer. As a result we get a complete (strong) alignment for
states with $J>1/2$ in the produced system. In the $^9$He case
only the magnetic substates with $M= \pm 1/2$ should be populated
for $J^{\pi}=5/2^+$, $3/2^-$, and $3/2^+$ states. This strongly
reduces possible ambiguity in the analysis of correlation
patterns. E.g.\ in the case of zero spin particles the zero
geometry experiments give very clean pictures with angular
distributions described by pure Legendre polynomial $|P^0_l|^2$.
The situation in the case of nonzero spin particles involved is
more complicated (see detailed discussion in Ref.\ \cite{gol05b}),
and diverse correlation patterns are possible.

We have found that the observed experimental picture (Figs.\ \ref{fig:expth},
\ref{fig:distrib}) can be well explained in a simple model
involving only three low-lying states: $1/2^+$, $1/2^-$, and $5/2^+$. The
inelastic cross in the DWBA ansatz is written as
\begin{eqnarray}
\frac{d\sigma(\Omega_{^9\text{He}})}{dE \,d \Omega_{^8\text{He}}} \sim 
\frac{v_{f}}{v_{i}}\,\sqrt{E} \,
\sum \nolimits_{MM_S} \left| \sum \nolimits_J\langle \Psi^{JMM_S}_f \left|
V \right| \Psi_i \rangle \right|^2  \label{eq:sigma}\\
= \frac{v_f}{v_i} \sum _{MM_S} \sum _{JJ'} \sum _{M'_lM_l}\! \rho_{JM}^{J'M}
C^{J'M'}_{l'M'_lSM_S} C^{JM}_{lM_lSM_S} \, Y^{\ast}_{lM'_l}
Y_{lM_l}\,.
\nonumber
\end{eqnarray}
For the density matrix the generic symmetries are
\[
\rho_{JM}^{J'M'}= \left(\rho^{JM}_{J'M'}\right)^* \quad ; \qquad
\rho_{JM}^{JM'}= (-)^{M+M'}\rho^{J-M}_{J-M'} \; ,
\]
and properties specific to coordinate choice (spirality representation) and
setup (zero geometry) are
\[
\rho_{JM}^{J'M'} \sim \; \delta_{M,M'}(\delta_{M,1/2}+\delta_{M,-1/2}) \; .
\]
For the density matrix parametrization we use the following model
for the transition matrix. The wave function (WF) $\Psi_f$ is
calculated in the $l$-dependent square well (with depth parameters
$V_l$). The well radius is taken $r_0=3$ fm, which is consistent
with typical R-matrix phenomenology $1.4 A^{1/3}$. The energy
dependence of the velocities $v_i$, $v_f$  (in the incoming $^8$He-$d$ and  
outgoing $^9$He-$p$ channels) and WF $\Psi_i$ is neglected for our range of 
$^9$He energies. The term $ V \left| \Psi_i \right. \rangle $, describing the 
reaction mechanism, is approximated by radial $\theta$-function:
\[
V \left| \Psi_i \right. \rangle \rightarrow C_l \;r^{-1}\,\theta(r_0-r) \;
[Y_{l}(\hat{r}) \otimes \chi_{S}]_{JM} \; ,
\]
where $C_l$ is (complex) coefficient defined by the reaction mechanism. For
$\bigl|\rho_{J\pm 1/2}^{J'\pm 1/2}\bigr|$ denoted as $A_{l'l}$ the cross section
as a function of energy $E$ and  $x=\cos(\theta_{^8\text{He}})$ is
\begin{eqnarray}
\frac{d\sigma(\Omega_{^9\text{He}})}{dE \;d x} \sim \frac{1}{\sqrt{E}}
\left[ \rule{0pt}{12pt} 4 A_{00} + 4 A_{11} + 3(1-2x^2+5x^4) A_{22}
\right. \nonumber \\
+ \left. 8 \, x \cos(\phi_{10})A_{10}  + 4 \sqrt{3} \, x(5x^2 - 3)
\cos(\phi_{12}) A_{12}  \right] \, . \,
\label{eq:sigma-full} \\
A_{l'l}=|A_{l'}| |A_{l}|\; , \;\;A_l= C_l
N_l(E)\,e^{i\delta_l(E)}\int_0^{r_0}dr j_l(q_lr)\; , \nonumber \\
 q_l=\sqrt{2M(E-V_l)}\; , \;\;
 \phi_{l'l}(E) = \phi^{(0)}_{l'l} + \delta_{l'}(E) - \delta_{l}(E)\, ,
\nonumber
\end{eqnarray}
where $N_l$ is defined by matching condition on the well boundary
for internal function $j_l(q_lr)$. The three coefficients $C_l$
give rise to the two phases $\phi^{(0)}_{10}$ and
$\phi^{(0)}_{12}$.

Positions and widths of the states are fixed by the three
parameters $V_l$. Their relative contributions to the inclusive
energy spectrum (Fig.\ \ref{fig:distrib}a--c) are fixed by the
three parameters $|C_l|$. The phase $\phi^{(0)}_{10}$ is fixed by
the angular distributions at low energy, where the contribution of
the $d$-wave resonance is small. After that the phase
$\phi^{(0)}_{12}$ was varied to fit the angular distributions at
higher energies (Fig.\ \ref{fig:distrib}, $E>2.2$ MeV). So, the
model does not have redundant parameters and the ambiguity of the
theoretical interpretation is defined by the quality of the data.

\begin{table}[b]
\caption{Parameters of theoretical model used in the work. $V_i$ values are in
MeV, weight coefficients for different states are normalized to unity $\sum
|C_i|^2=1$.}
\begin{ruledtabular}
\begin{tabular}[c]{lccccccc}
Set   & $|C_0|^2$ & $|C_1|^2$ & $V_0$ & $V_1$ & $V_2$ & $\phi^{(0)}_{10}$ &
$\phi^{(0)}_{12}$  \\
\hline
1  & 0.26 & 0.35 & $-4.0$   & $-20.7$ & $-43.4$ & $0.80 \pi$ & $-0.03 \pi$ \\
2  & 0.03 & 0.52 & $-4.0$   & $-20.7$ & $-43.4$ & $0.85 \pi$ & $-0.02 \pi$ \\
3  & 0.12 & 0.43 & $-5.817$ & $-20.7$ & $-43.4$ & $1.00 \pi$ & $-0.03 \pi$ \\
\end{tabular}
\end{ruledtabular}
\label{tab:th}
\end{table}

We have found that the weight and interaction strength for the
$1/2^+$ state can be varied in a relatively broad range, still
providing a good description of data. The results of MC
simulations of the experiment with different $s$-wave
contributions are shown in Fig.\ \ref{fig:distrib}a--c. The
parameter sets of the model are given in Table \ref{tab:th}; sets
1 and 2 correspond to small scattering length ($a=-4$ fm) and
different weights of $s$-wave (largest and lowest possible), set 3
has $a=-25$ fm and largest possible weight of $s$-wave. It can be
seen that the agreement with the data deteriorates when the
population of the $s$-wave continuum falls, say, below $15-25 \%$
of the $p$-wave. On the other hand the large negative scattering
length has a drastic effect below 0.5 MeV. The energy resolution
and the quality of the measured angular distributions are not
sufficient to draw solid conclusions about the exact properties of
the $s$-wave contribution. The situations with the large
contribution of the $s$-wave cross section but with moderate
scattering length (say $a
> -20$ fm) seem to be more plausible. Measurements with better
resolution are required to refine the properties of the $1/2^+$
continuum.

Position of the $d$-wave resonance is not well defined in our analysis of data
due to the efficiency fall in the high-energy side of the spectrum. This can be
well seen from the comparison of theoretical inputs and MC results in Fig.\
\ref{fig:distrib}a--c. The lower limit for the resonance energy of 4.2 MeV is in
a good agreement with the value 4.0 MeV found in \cite{gol03}. A broader energy
range measured for $^9$He is needed to resolve the $5/2^+$ state completely and
to make the
angular distribution analysis more restrictive.


\textit{Discussion.} --- It should be noted that the interference of any other
combination of $s$- $p$- $d$-wave states {\em can not} lead to the required 
forward-backward asymmetry in the whole energy range. The correlation terms 
[square brackets in Eq.\ (\ref{eq:sigma-full})] are
\begin{eqnarray}
\left[\rule{0pt}{9pt}\ldots \right] &= &2 A_{00} + 2 A_{11} + (1+3x^2) A_{22}
+ 4  x \cos(\phi_{10}) A_{10}
\nonumber \\
& +& 2 \sqrt{2} (3x^2 - 1) \cos(\phi_{20}) A_{20}\,  ,
\nonumber \\
\left[ \rule{0pt}{9pt}\ldots  \right] &=& 4 A_{00} + 2 (1+3x^2) A_{11} +
3(1-2x^2+5x^4) A_{22}  \,,
\nonumber \\
\left[ \rule{0pt}{9pt} \ldots \right] &= &2 A_{00} + (1+3x^2) (A_{11} +  A_{22})
+ 2 \, x (9x^2-5)
\nonumber \\
& \times & \cos(\phi_{12})A_{12} + 2 \sqrt{2}  (3x^2 - 1)
\cos(\phi_{20}) A_{20} \, ,
\nonumber
\end{eqnarray}
for $\{s_{1/2},p_{1/2},d_{3/2} \}$, $\{s_{1/2},p_{3/2},d_{5/2} \}$, and
$\{s_{1/2},p_{3/2},$ $d_{3/2} \}$ sets of states respectively. The
asymmetric term ($\sim x)$ is present here either for $s$-$p$ interference only
or for $p$-$d$ only or for neither.

\begin{figure}
\includegraphics[width=0.45\textwidth]{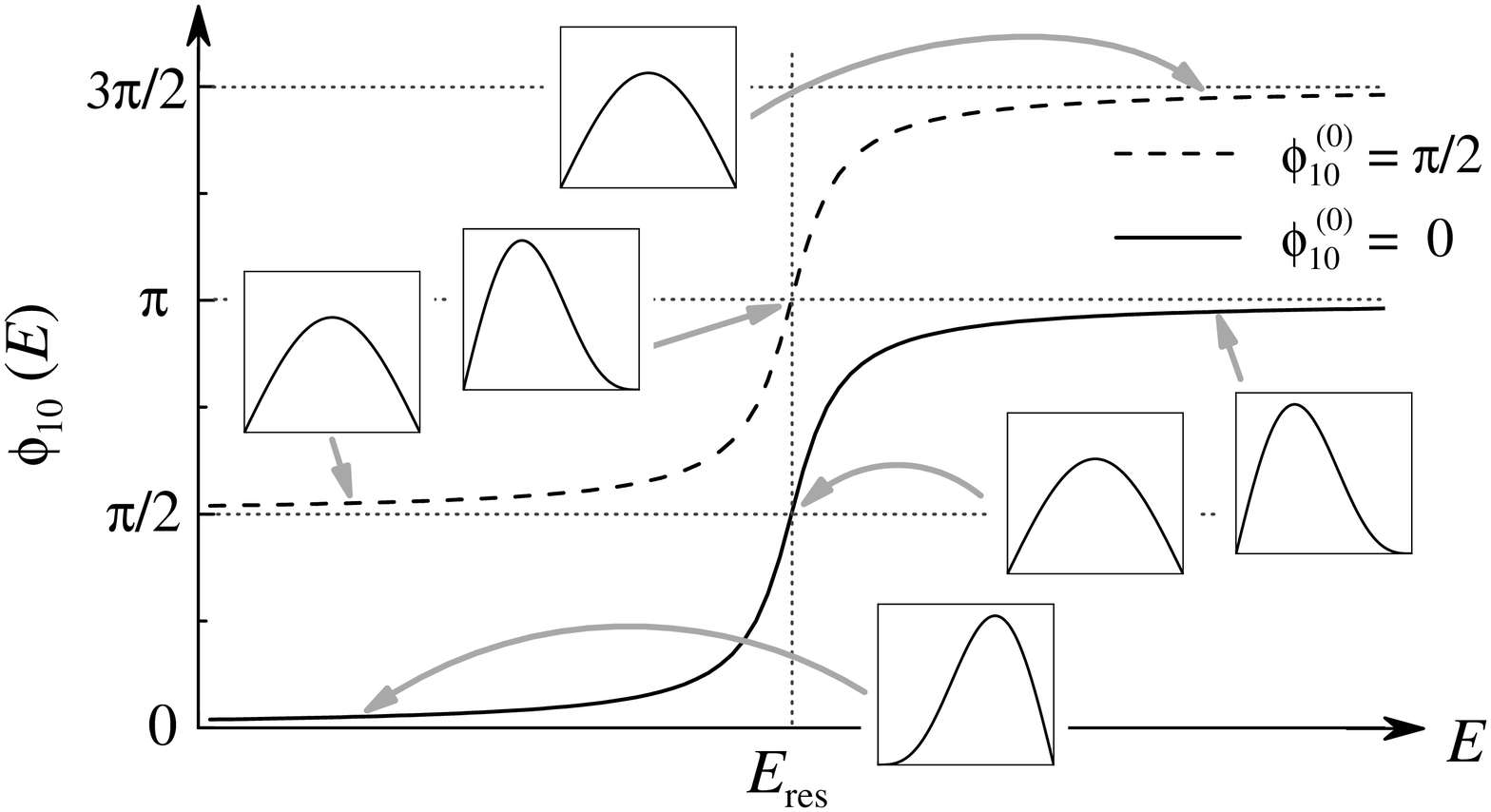}
\caption{Schematic illustration of possible behavior of angular
distributions (shown by inserts for angles $\theta_{^8\text{He}}$
from $0^{\circ}$ to $180^{\circ}$) due to $s_{1/2}$-$p_{12}$
interference around a narrow resonance for different phases
$\phi_{10}^{(0)}$.}
\label{fig:qualit}
\end{figure}

The angular distributions in energy bins (Fig.\ \ref{fig:distrib})
provide a good indication that a {\em narrow} $p_{1/2}$ state is
not populated in the reaction. Fig.\ \ref{fig:qualit} shows
qualitatively what happens with angular distribution if there is a
narrow resonance. The phase shift changes across the narrow
resonance to a value close to $\pi$ and the character of angular
distribution should change drastically within this energy range.
No trend of this kind is observed in Fig.\ \ref{fig:distrib}.
Phase shift for the {\em broad} $1/2^-$ state changes slowly and
hardly achieves $\pi /2$ in our calculations. This allows to
explain the smooth behaviour of asymmetry up to 3 MeV.

The existing experimental data can be regarded as not
contradicting our results. In Refs.\ \cite{set87,boh99} the narrow
states were observed with low statistics (15--40 events/state). If
the narrow states really exist they should be observable even at
such a low counting rates. However, simulations show that if the
cross section behavior is smooth, ``statistically driven'' narrow
structures are quite probable in such a situation. A look at the
data of Ref.\ \cite{gol03} also shows that states (which should
have analogues in $^9$He) at 2.2 MeV and 4.0 MeV are absolutely
evident in the data. However, the presence of a narrow 1.1 MeV
state is more likely not to contradict the data, rather than
necessarily follow from these.

The idea that only the $1/2^-$ resonance state can be found in the
low energy region not only looks natural, but also finds support
in the recent theoretical studies. In Ref. \cite{vol05}, which
deals with the whole chain of helium isotopes in continuum shell
model, the $1/2^-$ state is located at 1.6 MeV above the
$^8$He+$n$ threshold and the width is $\sim 0.6$ MeV indicating
the dominant single particle component in the WF. The $3/2^-$
state is predicted to be at 6.6 MeV and relatively narrow ($\sim
2.5$ MeV), what is natural for complicated particle-hole
excitations.



\textit{Conclusions.} --- We would like to emphasize the following
results of our study.

\noindent (i) Our data show two broad overlapping peaks (at 2 and
at 4.5 MeV) in the $^9$He spectrum. Statistics obtained in our
experiment is about factor of 10 higher than in the previous works
\cite{set87,boh99}. Our resolution is sufficient to resolve narrow
low-lying states. Even if a narrow $p_{1/2}$ is not resolved, the
rapid change of phase around resonance energy should produce the
change in the forward-backward asymmetry, which is also not seen
in the data.

\noindent (ii) An essential contribution of the $s$-wave $1/2^+$ state is
evident from the data. It is manifested in two ways: (a) Large
forward-backward asymmetry at $E \leq 3$ MeV and (b) accumulation
of counts around the threshold, which should not take place for
typical cross section behavior for higher $l$-values. A limit $a
> -20$ fm is obtained for the scattering length of this state.

\noindent (iii) The proposed spin assignment $\{s_{1/2},p_{1/2},d_{5/2} \}$ is
unique, as no other reasonable set of low-lying states can provide the observed
correlation pattern.

\noindent (iv) The experimental data are well described in a
simple single-particle potential model, involving only basic theoretical 
assumptions about the reaction mechanism and the low-energy spectrum of $^9$He. 
This supports the idea that $^8$He (having closed $p_{3/2}$ subshell) presents a 
``good'' core in the $^{9}$He structure.



\textit{Acknowledgments.} ---  This work was supported by the Russian
Foundation for Basic Research grants 02-02-16550, 02-02-16174, 05-02-16404, and
05-02-17535 by the INTAS grantS 03-51-4496 and 03-54-6545. LVG acknowledge the
financial support from the Royal Swedish Academy of Science and Russian Ministry
of Industry and Science grant NS-1885.2003.2.



\end{document}